\documentclass[12pt]{article}
\usepackage{amssymb}
\usepackage{amsfonts}
\usepackage{amsmath}
\usepackage{graphicx}
\usepackage{color}
\usepackage{rotating}
\usepackage{makecell}
\usepackage{braket}

\newcommand{\ve}{\varepsilon}

\newcommand{\+}{{\dagger}}

\newcommand{\pa}{\partial}

\newcommand{\beq}{\begin{equation}}
\newcommand{\eeq}{\end{equation}}
\newcommand{\eq}{\end{equation}}
\newcommand{\bea}{\begin{eqnarray}}
\newcommand{\eea}{\end{eqnarray}}

\renewcommand{\and}{{\quad{\rm and}\quad}}

\advance \textheight by \topskip
\makeatletter
\@addtoreset{equation}{section}

\makeatother
\begin{document}

\title{
{\bf  Confluent
Crum-Darboux transformations in Dirac Hamiltonians with $PT$-symmetric Bragg gratings }
}

\author{ 
{\small \textrm{\textup{\textsf{ 
Francisco Correa${}^{a,b}$, 
V\'{\i}t Jakubsk\'y${}^{c}$
}}} } \\
[15pt]
{\small \textit{ ${}^{a}$Leibniz Universit\"at Hannover, Appelstra\ss{}e 2, 30167 Hannover, Germany }}\\
{\small \textit{ ${}^{b}$Instituto de Ciencias F\'isicas y Matem\'aticas,
Universidad Austral de Chile, Casilla 567, Valdivia, Chile}}\\
{\small \textit{${}^{c}$Department of Theoretical Physics, 
Nuclear Physics Institute, 25068  \v Re\v z, 
Czech Republic}}\\
[10pt]
 \sl{\small{E-mails: 
francisco.correa@uach.cl, jakub@ujf.cas.cz}} 
}

\date{}

\maketitle

\begin{abstract}
We consider optical systems where propagation of light can be described by a Dirac-like equation with $PT$-symmetric Hamiltonian. In order to construct exactly solvable configurations, we extend the confluent Crum-Darboux transformation for the one-dimensional Dirac equation. The properties of the associated intertwining operators are discussed and the explicit form for higher-order transformations is presented. We utilize the results to derive a multi-parametric class of exactly solvable systems where the balanced gain and loss represented by the $PT$-symmetric refractive index can imply localization of the electric field in the material.
\end{abstract}

\section{Introduction}
In specific situations, the propagation of light in optical systems can be described by equations that are at home in quantum mechanics. Indeed, the Helmholtz equation in paraxial approximation acquires the form of a Schr\"odinger-like equation \cite{Milonni}. One deals with coupled differential equations of the Dirac-type within the context of the coupled-wave theory of distributed feedback lasers  \cite{kogelnik,Carrol}. The interaction present in the associated Hamiltonians depends on the optical properties of the material the light is propagating in. These optical characteristics can be described by the refractive index. It can be position-dependent in an optically inhomogeneous material, it can be also  complex valued when gain and loss occur in the system \cite{eo1,eo2,eo3}.

The description of optical systems with complex refractive index departs from the concept of standard quantum mechanics, where the operators are required to be hermitian. However, a link can be still established within the realm of $PT$-symmetric quantum mechanics, where the requirement of hermiticity is relaxed and replaced by $PT$-symmetry, where $P$ is the space inversion and $T$ is the time reversal operator \cite{BR,MR1,MR2,benderZnojilmostafazadeh1,benderZnojilmostafazadeh2,benderZnojilmostafazadeh3}.
The relation between optics and $PT$-symmetric quantum mechanics has been exploited extensively in the recent years. Optical systems described by a Schr\"odinger equation with complex $PT$-symmetric potential were analyzed e.g. in  \cite{ui1,ui2,ui3,ui4,i1,i2,bp1,bp3}.
Those described by Dirac equation were discussed in \cite{longhiPRL} where the attention was paid to the spectral singularities and the $PT$-symmetry breaking.

The light propagating through a non-uniform Bragg grating with small fluctuations of the refractive index can be well described by the coupled mode theory \cite{Carrol,Sipe}. The Maxwell equations 
for a monochromatic electromagnetic field of the form
\begin{align}\notag
\mathbf{E}=\mathbf{e_1} \left( E(x_3)e^{-i\omega t}+E^*(x_3)e^{i\omega t} \right) \, , \quad 
\mathbf{H}=\mathbf{e_2} \left( H(x_3)e^{-i\omega t}+H^*(x_3) e^{i\omega t} \right) \, .
\end{align} 
reduce to 
$ \frac{\partial}{\partial x_3} E(x_3)=i\omega\mu_0 H(x_3)$, $\frac{\partial}{\partial x_3}H=i\omega \epsilon_0 n^2(x_3)E(x_3).$ Combining the two equations, we find that the electric field has to satisfy
\begin{equation}\label{E2}
\frac{\partial^2}{\partial x_3^2}E(x_3)+k^2\left(\frac{n(x_3)}{n_0}\right)^2 E(x_3)=0.
\end{equation}
Here, $k=\omega n_0/c$, $\mu_0\epsilon_0=c^{-1}$, and $n_0$ is the reference refractive index. Fixing the electric field in the form of two counter-propagating waves,
\begin{equation}
E(x)=u(x)\exp\left(ix+\frac{i}{2}\phi(x)\right)+v(x)\exp\left(-ix-\frac{i}{2}\phi(x)\right),\quad x=k_0 x_3,
\end{equation}
where $k_0$ is defined as the wave number of light at the Bragg scattering resonance frequency $\omega_0=ck_0/n_0$, the equation (\ref{E2}) can be brought into the form of two coupled equations \cite{Sipe},
\begin{align}\label{coupledoptics}
\partial_{x}u=i(\rho(x)u(x)+\kappa(x)v(x)),\quad
\partial_{x}v=-i(\rho(x)v(x)+\kappa(x)u(x)),
\end{align}
where $\rho(x)=\sigma(\xi)+\Delta-\frac{1}{2}\partial_{x}\phi(x)$ and  $\Delta=(\omega-\omega_0)/\omega_0$.  The functions $\sigma(x)$, $\kappa(x)$ determine the profile of the refractive index 
$n(z)=n_0(1+\sigma(x)+2\kappa(x)\cos(2x+\phi(x)))$. In order to keep fluctuations of $n(x)$ small, we require $|\sigma(x)|<<1$ and $|\kappa|<<1$.
With the use of $F=(u,v)^T$, we can rewrite (\ref{coupledoptics}) as
\begin{align}\label{DE}
HF(x)=(-i\sigma_3\partial_{x}+i\sigma_2\rho(x)-\sigma_1\kappa(x))F(x)=0,
\end{align}
which coincides with the one-dimensional \emph{non-hermitian} stationary Dirac equation at zero-energy \footnote{Using the ansatz $E(x,t)=u(x,t)\exp\left(ix+\frac{i}{2}\phi(x)\right)+v(x,t)\exp\left(-ix-\frac{i}{2}\phi(x)\right)$, there would appear $i\partial_tF$ on the right hand side of (\ref{DE}), see \cite{Carrol}. Let us notice that Dirac-like equation appears also in description of another optical system where two identical coupled $PT$ wave guides are considered, see \cite{po1}.}. The systems described by  Eq. (\ref{DE}) will be of our main interest.

The Crum-Darboux transformation is a differential operator that annihilates fixed eigenstates of the initial, known\footnote{The solutions of the associated initial stationary equation are supposed to be known.}, Hamiltonian \cite{CKSsusy,Matveev}. There exists an unambiguous prescription for the new Hamiltonian such that it is intertwined with the initial one by this transformation. The intertwining relation implies that the exact solutions for the stationary equation of the new system are obtained by acting on the eigenstates of the initial Hamiltonian with the intertwining operator.  
Besides being very fruitful for quantum mechanics, see e.g. \cite{AndrianovCannata} for a review, it proves to be an effective tool for the analysis of optical systems \cite{optsusy1,optsusy3,optsusy4,susyPTopt4,susyPTopt6}.

When the Crum-Darboux transformation annihilates eigenstates and Jordan states associated with a single eigenvalue is called \textit{confluent}. In the literature, it was discussed mostly for Schr\"odinger Hamiltonians in the context of quantum systems \cite{BICSukhatme,Pursey,CSUSY1,CSUSY4,CSUSY5,CSUSY6,Carinena:2016hfq}, as well as for optical settings with complex refractive index \cite{susyPTopt3,CJPCSUSY}. Recently, it was also used for the construction of a limited class of Dirac Hamiltonians in \cite{AxelCDirac}.

In the current article,  we will find $\rho(x)$ and $\kappa(x)$ with the help of Crum-Darboux transformations such that (\ref{DE}) is exactly solvable. We modify the confluent Crum-Darboux transformation for the use with Dirac Hamiltonians with a generic potential. 
We consider chains of confluent Crum-Darboux transformations and show explicit formulas for both the higher-order intertwining operator and the new Hamiltonian. In particular, we focus on Crum-Darboux transformations that render the new Hamiltonian $PT$-symmetric. Finally, we construct a class of $PT$-symmetric Dirac operators whose relevance for description of optical systems is discussed.

\section{First-order transformations and their chains}
In order to explain in more details the confluent Crum-Darboux transformations, let us start with the construction of a first order transformation. 
We set the initial, one-dimensional Dirac Hamiltonian in the following form
\begin{align}\label{hgen}
H_0=-i\sigma_3\partial_x+V_0,
\end{align} 
where $V_0$ is an arbitrary matrix function. Following \cite{Samsonov}, we make the following ansatz for the intertwining operator
\begin{equation}\label{L}
 L_1=\partial_x-U_xU^{-1}.
\end{equation}
Since we assume $U$ is an invertible matrix and $U_x\equiv \partial_x U$,  by construction, the above equation means that $L_1 U=0$. We want $L_1$ to be an intertwining operator between $H_0$ and another Hamiltonian $H_1=-i\sigma_3\partial_x+V_1$. Hence, the explicit form of both $L_1$ and $V_1$ should be fixed such that there holds
\begin{equation}
L_1H_0=H_1 L_1\, .
\end{equation}
Comparing the coefficients at the corresponding derivatives, we get two equations for the unknown matrices $V_1$ and $U$,
\begin{align}\label{V_1}
 V_1=V_0-i [U_xU^{-1},\sigma_3], \quad \left( U^{-1}H_0U\right)_x=0.
\end{align}
The first one fixes $V_1$. The second one 
is a differential equation for $U$. It is satisfied whenever $H_0U=U\Lambda$ with  $\Lambda$ being a constant matrix. As any matrix with complex elements can be transformed into a Jordan form,  we can fix $U$ in such way that
\begin{equation}\label{genSams}
 H_0U=U\Lambda,\quad\text{for}\quad \Lambda=\left(\begin{array}{cc}\lambda_1&0\\0&\lambda_2\end{array}\right)\quad \text{or}\quad \Lambda=\left(\begin{array}{cc}\lambda&1\\0&\lambda\end{array}\right).
\end{equation}
Hence, we can write down $U$ as  
\begin{equation}
U=(\Omega_1,\Omega_2).
\end{equation}
The spinors $\Omega_1$ and $\Omega_2$ either satisfy $H_0\Omega_a=\lambda_a\Omega_a$, in the case when $\Lambda$ is diagonal\footnote{Let us notice that $H_1$ reduces to $H_0$ when $\lambda_1=\lambda_2$.}, or $(H_0-\lambda)\Omega_1=0$ and $(H_0-\lambda)\Omega_2=\Omega_1$ when $\Lambda$ has the form of a Jordan block. The latter choice represents a direct generalization of \cite{Samsonov}, where only a diagonal $\Lambda$  was considered. 

Once $U$ is fixed, the intertwining operator $L_1$ as well as $H_1$ are uniquely defined. The operator (\ref{L}) annihilates both vectors $\Omega_a$, $L\Omega_a=0$, $a=1,2$. It is worth noticing that, at the moment, the present construction is formal and an additional care is needed to obtain a physically relevant result. In particular, the requirement of regularity has to be imposed on the intertwining operator such that the new Hamiltonian $H_1$ does not contain any new singularities. This is equivalent to the requirement that $\det U\neq0$ for the considered domain of $x$. We will discuss this issue later on.

\begin{figure}[h!]
	\centering
	\includegraphics[scale=0.35]{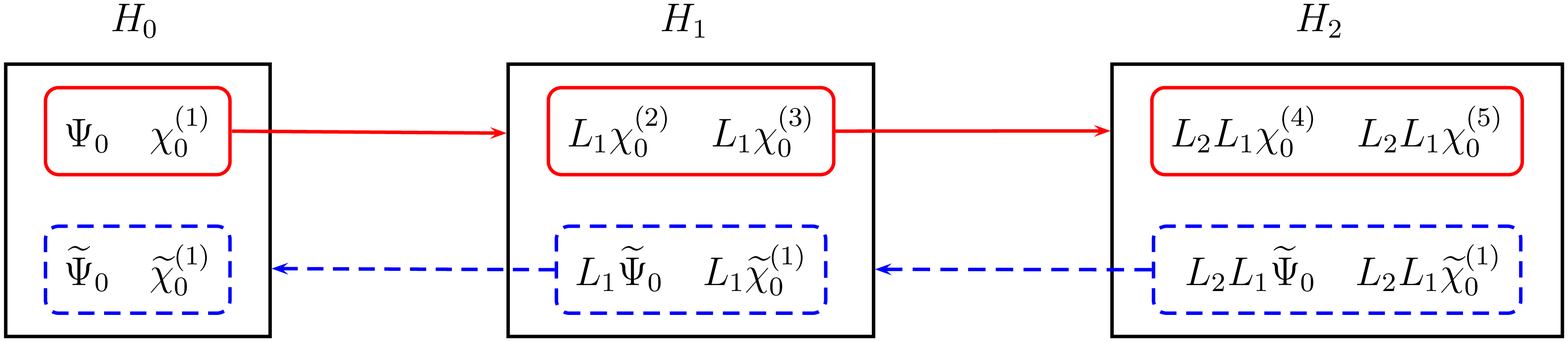}
	\caption{Schematically illustrated action of $L_1$, $L_2$ (red arrows) and $L^\sharp_1$, $L_{2}^\sharp$ (dashed blue arrows) on the eigenvectors of the associated Jordan states of $H_0$ (left box), $H_1$ (center), and $H_2$ (right).}
	\label{polyhedra0}
\end{figure}

Let us continue our discussion with the case of a chain of consecutive first order Crum-Darboux transformations. We focus on the situation where the intertwining operators are systematically constructed from the Jordan states associated with a fixed eigenvalue $\lambda_*$. 
First, we shall fix the notation. Given a Hamiltonian $H_0$, we denote the two independent (not necessarily physical) eigenvectors of $H_0$ corresponding to an eigenvalue $\lambda$ as  $\Psi_0$ and $\widetilde{\Psi}_0$,
\begin{align}
(H_0-\lambda)\Psi_0=0,\quad (H_0-\lambda)\widetilde{\Psi}_0=0.
\end{align}
The Jordan states ${\chi}^{(n)}_0$ and $\widetilde{\chi}^{(n)}_0$ associated with $\Psi_0$ and $\widetilde{\Psi}_0$, respectively,  can be defined as the solutions of the $(n+1)$th iterated Dirac equation, $(H_0-\lambda)^{n+1} {\chi}^{(n)}_0=(H_0-\lambda)^{n+1} \widetilde{\chi}^{(n)}_0=0$, in the following form \cite{CSUSY6},
\begin{align}\label{JS1}
\chi^{(n)}_0=\frac{\partial^n \Psi_0 }{\partial \lambda^n }+\sum_{k=0}^{n-1}(c_k^{(n)}\chi^{(k)}_0+d_k^{(n)}\tilde{\chi}^{(k)}_0),\quad  \quad (H_0-\lambda)^n \chi^{(n)}_0 =\Psi_0 \, ,\\
\label{JS2}
\widetilde{\chi}^{(n)}_0=\frac{\partial^n \widetilde{\Psi} }{\partial \lambda^n }+\sum_{k=0}^{n-1}(\tilde{c}_k^{(n)}\chi^{(k)}_0+\tilde{d}_k^{(n)}\tilde{\chi}^{(k)}_0),\quad  \quad (H_0-\lambda)^n \widetilde{\chi}^{(n)}_0 =\widetilde{\Psi}_0 \, ,
\end{align}
where $c_k^{(n)}$, $\tilde{c}_k^{(n)}$, $d_k^{(n)}$ and $\tilde{d}_k^{(n)}$ are some complex numbers and ${\chi}^{(0)}_0=\Psi_0$ and $\widetilde{\chi}^{(0)}_0=\widetilde{\Psi}_0$.
In particular, the Jordan states satisfy
\begin{align}\label{bla}
(H_0-\lambda) {\chi}^{(n)}_0={\chi}^{(n-1)}_0, \quad (H_0-\lambda) \widetilde{\chi}^{(n)}_0=\widetilde{\chi}^{(n-1)}_0.
\end{align}
As the first link in the chain of transformations we are going to discuss, we define $L_1$ such that it annihilates $\Psi_0$ and ${\chi}_0^{(1)}$, $(H_0-\lambda_*)\Psi_0=(H_0-\lambda_*){\chi}_0^{(1)}-\Psi_0=0$  and intertwines $H_0$ with the Hamiltonian $H_1$,
\begin{align}\label{actionint}
L_1\Psi_0=L_1{\chi}^{(1)}_0=0,   \quad L_1H_0=H_1L_1\, .
\end{align}
The eigenstates $\Psi_1$ and $\tilde{\Psi}_1$ of $H_1$, $(H_1-\lambda)\Psi_1=(H_1-\lambda)\tilde{\Psi}_1=0$, can be written for any $\lambda$ as
\begin{align}\label{JSS1}
\widetilde{\Psi}_1=L_1\widetilde{\Psi}_0,\quad \Psi_1 =
\begin{cases}
L_1\Psi_0  \quad \text{for} \quad \lambda\neq \lambda_* \\
L_1 {\chi}^{(2)}_0  \quad \text{for} \quad \lambda=\lambda_*. \\
\end{cases}
\end{align}
Here, we have utilized (\ref{bla}) and (\ref{actionint}) which imply
$L_1{\chi}^{(1)}_0=L_1(H_0-\lambda_*) {\chi}^{(2)}_0= (H_1-\lambda_*) L_1{\chi}^{(2)}_0=0$.
The Jordan states $\chi_1^{(1)}$ or $\widetilde{\chi}_1^{(1)}$ of $H_1$ associated with either $\Psi_1$ or $\widetilde{\Psi}_1$, $(H_1-\lambda)\chi_1^{(1)}=\Psi_1$, $(H_1-\lambda)\tilde{\chi}_1^{(1)}=\tilde{\Psi}_1$, are
\begin{align}\label{JSS2}
\widetilde{\chi}^{(1)}_1=L_1\widetilde{\chi}^{(1)}_0,\quad \chi_1^{(1)} =
\begin{cases}
L_1\chi_0^{(1)} \quad \text{for} \quad \lambda\neq \lambda_*, \\
L_1 {\chi}^{(3)}_0  \quad \text{for} \quad \lambda=\lambda_*. \\
\end{cases}
\end{align}
For the second link of the chain, we fix $L_2$ such that $L_2\Psi_1=L_2\chi_1^{(1)}=0$, $(H_1-\lambda_*)\Psi_1=0$, $(H_1-\lambda_*)\chi_1^{(1)}=\Psi_1$. Then there exists $H_2$ such that the following relations hold true,
\begin{align}
L_1H_0=H_1L_1,\quad L_2H_1=H_2L_2,\quad L_2L_1H_0=H_1L_2L_1.
\end{align}
We can identify the chain of the two Darboux transformations with a second-order intertwining operator $\mathcal{L}_2=L_2 L_1$ that directly intertwines $H_0$ with $H_2$. $\mathcal{L}_2$ can be defined directly as the operator that annihilates $\Psi_0,\ {\chi}^{(1)}_0,\ {\chi}^{(2)}_0,\ {\chi}^{(3)}_0$, see (\ref{JSS1}) and (\ref{JSS2}), and the coefficient at the highest derivative is normalized to one. The chain is illustrated in Fig. \ref{polyhedra0}. 

The Hamiltonians $H_0$, $H_1$ are not hermitian in general. Hence, the construction of the intertwining operator $L_1^{\sharp}$ which satisfies $H_0L_1^{\sharp}=L_1^{\sharp}H_1$ is less straightforward than in the hermitian case. Let us define $L_1^{\sharp}=\partial_x-U^{\sharp}_x(U^{\sharp})^{-1}$ where $U^{\sharp}=(\widetilde{\Psi}_1,\widetilde{\chi}^{(1)}_1)$. Then $L_1^{\sharp}$ intertwines $H_1$ and $H_0$,
\begin{equation}\label{intLsharp}
H_0L_1^{\sharp}=L_1^{\sharp}H_1,\quad L_1^{\sharp}\widetilde{\Psi}_1=L^{\sharp}_1\widetilde{\chi}_1^{(1)}=0.
\end{equation}
To prove this relation, notice first that 
$L_1^{\sharp}L_1=(H_0-\lambda_*)^2$
as the operators have the same kernel and the same coefficient at the highest derivative. Then, we can rewrite $[L_1^{\sharp}L_1,H_0]=0$ as $(L_1^{\sharp}H_1-H_0L_1^\sharp)L_1=0$. Hence, the relation (\ref{intLsharp}) is valid on the space of all eigenstates and associated Jordan states of $H_1$. The operator $L_1^\sharp$ acts like the ``inverse" operator to $L_1$, see Fig. \ref{polyhedra0} for illustration.

\section{Intertwining operators of higher order}

One can construct a chain of Darboux transformations of arbitrary length $n$ and identify it with a higher-order intertwining operator $\mathcal{L}_n$ such that 
\begin{align}\label{gnin}
{\cal L}_n H_0 = H_n {\cal L}_n, \qquad {\cal L}_n\Psi_0= {\cal L}_n{\chi}^{(j)}_0=0,\quad j=1,2,...,n-1.
\end{align}
We are going to write down the explicit form of $H_n$ and $\mathcal{L}_n$. Let us fix the initial Hamiltonian $H_0$,
\begin{align}\label{hbdg}
H_0=
\left(
\begin{array}{cc}
-i\pa_x&q_0  \\
 r_0 & i\pa_x
\end{array}
\right)\,,
\end{align} 
where $q_0(x)$ and $r_0(x)$ are arbitrary complex functions,
and a set of $2n$ spinors,
\begin{align}\label{set}
S_{2n}=\left\{ \Xi_1,\Xi_2, \ldots, \Xi_{2n-1},\Xi_{2n}\right\}, \qquad \Xi_i=\left( 
\begin{array}{c}
    f_i \\ 
    g_i \\ 
  \end{array}
\right).
\end{align}
We suppose that they represent a sequence of an eigenstate and its associated Jordan states corresponding to a single eigenvalue $\lambda_*$, 
$S_{2n}=\left\{ \Psi_0, {\chi}^{(1)}_0, {\chi}^{(2)}_0, \ldots, {\chi}^{(2n-1)}_0 \right\}, $ $(H_0-\lambda_*)\Psi_0=0$.
Then we suggest the following definitions of the intertwining operator ${\cal L}_n$, which annihilates the complete set (\ref{set}), 
\begin{align}\label{intertn}
{\cal L}_n= \left( 
\begin{array}{cc}
   \pa_x^n + \sum_{\ell=1}^{n-1} \frac{\det A[f^{(\ell)}]}{\det W}\pa_x^\ell &   \sum_{\ell=1}^{n-1} \frac{\det A[g^{(\ell)}]}{\det W}\pa_x^\ell \\
\sum_{\ell=1}^{n-1} \frac{\det B[f^{(\ell)}]}{\det W}\pa_x^\ell &  \pa_x^n + \sum_{\ell=1}^{n-1} \frac{\det B[g^{(\ell)}]}{\det W}\pa_x^\ell \\ 
  \end{array}
\right)\, ,\quad {\cal L}_n\Xi_j=0,\quad j=1,2,...,2n.
\end{align}
Here, we denoted
\begin{align}\label{wrons}
A= \left(
\begin{array}{cccccccc}
f^{(n)}  &    f^{(n-1)}  &  \ldots & f &  g ^{(n-1)} &   \ldots &g ^{\, \prime} & g \\
f _{1}^{(n)}  &    f _{1}^{(n-1)}  &  \ldots & f _{1} &  g _{1}^{(n-1)} &   \ldots &g _{1}^{\, \prime} & g _{1}  \\
f _{2}^{(n)}   &  f _{2}^{(n-1)}   &  \ldots & f _{2} &  g _{2}^{(n-1)} &   \ldots &g _{2}^{\, \prime} & g _{2}  \\ 
\vdots & \vdots &  &  & &  &\\
f _{2n}^{(n)}  &    f _{2n}^{(n-1)}  &  \ldots & f _{2n} &  g _{2n}^{(n-1)} &   \ldots &g _{2n}^{\, \prime} & g _{2n}  \\
\end{array}
\right) ,
\end{align}
\begin{align}\label{wrons}
B= \left(
\begin{array}{cccccccc}
g ^{(n)}  &    f ^{(n-1)}  &  \ldots & f  &  g ^{(n-1)} &   \ldots &g ^{\, \prime} & g   \\
g _{1}^{(n)}  &    f _{1}^{(n-1)}  &  \ldots & f _{1} &  g _{1}^{(n-1)} &   \ldots &g _{1}^{\, \prime} & g _{1}  \\
g _{2}^{(n)}   &  f _{2}^{(n-1)}   &  \ldots & f _{2} &  g _{2}^{(n)} &   \ldots &g _{2}^{\, \prime} & g _{2}  \\ 
\vdots & \vdots &  &  & &  &\\
g _{2n}^{(n)}  &    f _{2n}^{(n-1)}  &  \ldots & f _{2n} &  g _{2n}^{(n)} &   \ldots &g _{2n}^{\, \prime} & g _{2n}  \\
\end{array}
\right).
\end{align}
The matrix $A[h^{(i)}]$ ($B[h^{(i)}]$) in (\ref{intertn}) is obtained from $A$ ($B$) by substituting all the entries in the first row by zeros \emph{except} the position $h^{(i)}$ where $h^{(i)} \rightarrow 1$. Here, $h$ is either $f$ or $g$. To get $A[f^{(1)}]$ in the case $n=2$, we have to substitute the first row $\left[  f^{\prime\prime}  ,  f^{\prime} , f ,  g^{\prime} , g \right]$ in the matrix $A$ by $ \left[0,1,0,0,0\right]$.
The matrix $W$ corresponds to the generalized Wronskian of the set (\ref{set}),
\begin{align}\label{wrons}
W= \left(
\begin{array}{cccccccc}
 f _{1}^{(n-1)}  &    f _{1}^{(n)}  &  \ldots & f _{1} &  g _{1}^{(n-1)} &   \ldots &g _{1}^{\, \prime} & g _{1}  \\
f _{2}^{(n-1)}   &  f _{2}^{(n)}   &  \ldots & f _{2} &  g _{2}^{(n-1)} &   \ldots &g _{2}^{\, \prime} & g _{2}  \\ 
 \vdots & \vdots &  &  & &  &\\
 f _{2n}^{(n-1)}  &    f _{2n}^{(n)}  &  \ldots & f _{2n} &  g _{2n}^{(n-1)} &   \ldots &g _{2n}^{\, \prime} & g _{2n}  \\
\end{array}
\right) .
\end{align}
The Hamiltonian $H_n$, defined in Eq. (\ref{gnin}), acquire the following form
\begin{align}\label{htilde}
H_{n}=
\left(
\begin{array}{cc}
-i\frac{d}{dx}& q_{n} \\
r_{n}  & i\frac{d}{dx}
\end{array} \right)\,,
\end{align}
where
\begin{align}\label{genrqn}
q_{n}= q+2i\frac{ \det D_q }{\det W }\, ,  \quad r_{n}=r+2i\frac{ \det D_r}{\det W },
\end{align}
and
\begin{align}\label{Dq}
D_q= \left(
\begin{array}{cccccccc}
g_{1}^{(n-2)}  &    g_{1}^{(n-3)}  &  \ldots & g_{1} &  f_{1}^{(n)} &   \ldots &f_{1}^{\, \prime} & f _{1}  \\
g _{2}^{(n-2)}   &  g _{2}^{(n-3)}   &  \ldots & g_{2} &  f_{2}^{(n)} &   \ldots &f_{2}^{\, \prime} & f_{2}  \\ 
\vdots & \vdots &  &  & &  &\\
g_{2n}^{(n-2)}  &   g_{2n}^{(n-3)}  &  \ldots & g_{2n} &  f _{2n}^{(n)} &   \ldots &f_{2n}^{\, \prime} & f_{2n}  \\
\end{array}
\right) ,
\end{align}
\begin{align}\label{Dr}
D_r= \left(
\begin{array}{cccccccc}
g_{1}^{(n)}  &    g_{1}^{(n-1)}  &  \ldots & g_{1} &  f_{1}^{(n-2)} &   \ldots &f_{1}^{\, \prime} & f _{1}  \\
g _{2}^{(n)}   &  g _{2}^{(n-1)}   &  \ldots & g_{2} &  f_{2}^{(n-2)} &   \ldots &f_{2}^{\, \prime} & f_{2}  \\ 
\vdots & \vdots &  &  & &  &\\
g_{2n}^{(n)}  &   g_{2n}^{(n-1)}  &  \ldots & g_{2n} &  f _{2n}^{(n-2)} &   \ldots &f_{2n}^{\, \prime} & f_{2n}  \\
\end{array}
\right).
\end{align}
Then the operators $H$, $H_n$ and $\mathcal{L}_n$ as defined (\ref{hbdg}) and (\ref{intertn})-(\ref{Dr}) satisfy the intertwining relation
\begin{equation}\label{nintertwining}
\mathcal{L}_n H= H_{n}\mathcal{L}_n.
\end{equation}
The formulas (\ref{intertn})-(\ref{Dr}) coincide with the those that appear in the literature for the usual Crum-Darboux transformations, e.g. \cite{Yurov}.

In the end of the section, let us comment on the construction of the inverse intertwining operator for the operator (\ref{hbdg}) and (\ref{htilde}).
The Hamiltonians (\ref{hbdg}) and (\ref{htilde}) satisfy the following  symmetry relation
\begin{align}
H_0=\sigma_1 H_0^{t} \sigma_1,\quad H_n=\sigma_1 H_n^{t} \sigma_1\, ,
\end{align}
where ${}^t$ stands for transposition. Transposing and multiplying (\ref{nintertwining}) by $\sigma_1$ from both sides, we get $\sigma_1 (H_{n}\mathcal{L}_n)^t \sigma_1  =\sigma_1 (\mathcal{L}_n H)^t \sigma_1 $ that can be written as
\begin{align}
(\sigma_1 \mathcal{L}_n^t \sigma_1)  H_{n} &=   H_0  (\sigma_1 \mathcal{L}_n^t \sigma_1)\, ,
\end{align}
which is the inverse intertwining relation for $H_0$ and $H_n$.

\section{$PT$-symmetric Dirac operators}
In order to simulate a balanced gain and loss in the system, we require the Hamiltonian $H_n$ to be $PT$-symmetric, where $P$ is the spatial inversion ($PxP=-x$) and $T$ is the time reversal  operator ($TxT=x$, $TiT=-i$). Let us suppose that the Hamiltonian $H_0$ is $PT$-symmetric, i.e. 
$$[H_0,PT]=0,$$ 
that implies 
$ [PT,q_0]=[PT,r_0]=0.$
Both eigenstates $\Psi$, $\widetilde{\Psi}$ and the associated Jordan states $\chi_0^{(n)}$, $\tilde{\chi}_0^{(n)}$ of $H_0$ can be fixed in such a way they have definite parity $\epsilon$ with respect to the operator $PT$,
\begin{eqnarray}
 &&PT\Psi_0=\epsilon\Psi_0,\quad PT\widetilde{\Psi}_0=-\epsilon\widetilde{\Psi}_0,\\
 &&PT\chi^{(n)}_0=\epsilon\, \chi^{(n)}_0,\quad PT\widetilde{\chi}^{(n)}_0=-\epsilon\, \widetilde{\chi}^{(n)}_0,\quad \epsilon^2=1. 
\end{eqnarray}
To satisfy these relations, the Jordan states (\ref{JS1}) and (\ref{JS2}) are fixed as 
\begin{align}\label{JS1PT}
\chi^{(n)}_0=\frac{\partial^n \Psi_0 }{\partial \lambda^n }+\sum_{k=0}^{n-1}(c_k^{(n)}\chi^{(k)}_0+i\, d_k^{(n)}\widetilde{\chi}^{(k)}_0),\quad  
\widetilde{\chi}^{(n)}_0=\frac{\partial^n \widetilde{\Psi}_0 }{\partial \lambda^n }+\sum_{k=0}^{n-1}(i\, \tilde{c}_k^{(n)}\chi^{(k)}_0+\tilde{d}_k^{(n)}\widetilde{\chi}^{(k)}_0), \, 
\end{align}
where the constants $c_k^{(n)}$, $\tilde{c}_k^{(n)}$, $d_k^{(n)}$ and $\tilde{d}_k^{(n)}$ must acquire real values.

We chose the set of seed, $PT$-symmetric, functions given by
$S_{2n}=\left\{ \Psi_0, {\chi}^{(1)}_0, {\chi}^{(2)}_0, \ldots, {\chi}^{(2n-1)}_0 \right\}$.
Using directly the formulas (\ref{wrons}), (\ref{Dq}), and (\ref{Dr}) we can see that
\begin{align}
[PT , \det W]=0, \quad \{ PT, \det D_q\}=0, \quad \{ PT, \det D_q\}=0\, ,
\end{align}
which imply that
$ [PT,q_n]=[PT,r_n]=0$.
As there also holds
 $PT \det A[\psi^{(i)}]PT=(-1)^{n-i}\det A[\psi^{(i)}],$
the Hamiltonian $H_n$ is therefore $PT$-symmetric,
\begin{equation}
[H_n,PT]=0.
\end{equation}
The intertwining operator $\mathcal{L}_n$ either commutes or anticommutes with $PT$, dependently on the value of $n$. For any $\Psi$, $PT\Psi=\epsilon\Psi$, we have
 $PT \mathcal{L}_n\Psi=\epsilon (-1)^n  \mathcal{L}_n\Psi$.

It is desirable that the potential terms of $H_n$, more specifically $q_n$ and $r_n$, does not display any new singularities. It is the Wronskian determinant, $\det W$, which encodes singularities of both $q_n$ and $r_n$. Thus, when we require the Hamiltonian $H_n$ to be regular, the Wronskian determinant should be nodeless. Let us take $n=1$ and
fix $\chi_0^{(1)}=\partial_\lambda\Psi_0+i d_{0}^{(1)}\widetilde{\Psi}_0$ where $d_{0}^{(1)} \in \mathbb{R}$, see (\ref{JS1PT}). Then we have 
\begin{equation}\label{detW}
 \det W[\Psi_0,\chi_0^{(1)}]=\det W[\Psi_0,\partial_\lambda\Psi_0]+i d_{0}^{(1)} \det W[\Psi_0,\widetilde{\Psi}_0].
\end{equation}
The second term is a nonzero number ($\Psi_0$ and $\tilde{\Psi}_0$ are the two independent solutions of the stationary equation). As we know that $PT\det WPT=\det W$, the number has to be purely purely real. It implies that the imaginary part of (\ref{detW}) is independent of $d_{0}^{(1)}$, whereas the real part of (\ref{detW}) contains $d_{0}^{(1)}$ as an additive constant. Hence, one can always fix $d_{0}^{(1)}$ such that the zeros of the real and imaginary parts of (\ref{detW}) are mismatched, giving rise to a singularity-free determinant $\det W$.

\section{Intensity of the electric field in $PT$-symmetric gratings: Examples}
Let us apply here the confluent Crum-Darboux transformation on an explicit Hamiltonian $H_0$. 

We identify the following initial Dirac Hamiltonian that, in quantum mechanics, would correspond to a free particle with possibly non-vanishing mass,
\begin{align}\label{free}
H_0=-i\sigma_3 \partial_x+\sigma_1 \delta=
\left(
\begin{array}{cc}
-i \pa_x & \delta \\
\delta & i \pa_x
\end{array}
\right)\,,\quad \delta \geq 0.
\end{align} 
Its spectrum\footnote{Here, we borrow terminology of quantum mechanics. The eigenfunctions associated with the allowed eigenvalues are either square integrable or correspond to the quantum mechanical scattering states. } consists of two bands of negative and positive energies $\ve\in(-\infty,-\delta] \cup [\delta,\infty)$, divided by a energy gap of magnitude $2\delta$. 
When comparing $H_1$ with the Hamiltonian in (\ref{DE}), we find that $\rho=\frac{q_1-r_1}{2}$, $\kappa=-\frac{q_1+r_1}{2}$, i.e.
\begin{equation}\label{hde}
H_1=-i\sigma_3\partial_x+\left(
\begin{array}{cc}
0 & q_{1} \\
r_{1}  &0 
\end{array} \right)=-i\sigma_3\partial_x+\frac{q_1-r_1}{2}\sigma_2+\frac{q_1+r_1}{2}\sigma_1.
\end{equation}
The eigenstates of $H_1$ can be obtained via (\ref{JSS1}). However, we are concerned only with the zero modes of $H_1$ as their components represent the intensities of electric field of the counter-running light in the grating.

The zero modes of $H_1$ differ qualitatively in dependence on the spectral gap of $H_1$ and the utilized transformation, based on the chosen seed states. When there is a gap, the two solutions of (\ref{DE})  are either expanding  exponentially towards infinity or one of them can be square integrable. If there is no gap, the zero modes correspond to the threshold of the continuum spectrum. One of them can be a bounded function. Let us elaborate these two cases below.

\subsection{Exponentially decaying electric field}
For $\delta>0$, it is convenient to parametrize the eigenfunctions and eigenvalues of $H_0$ in the following manner,
\begin{align}\label{eigenh}
\Psi_\theta
=\left(
\begin{array}{c}
\cosh \left(\delta  \sin\theta \, x +i \mu \right) \\
\cosh \left(\delta  \sin\theta \, x +i \theta +i \mu \right)\\
\end{array}
\right), \quad \widetilde{\Psi}_{\theta}=\left(
\begin{array}{c}
\sinh \left(\delta  \sin\theta \, x +i \mu \right) \\
\sinh \left(\delta  \sin\theta \, x +i \theta +i \mu \right)\\
\end{array}
\right),
\end{align}
\begin{align}\label{d}
H_0 \Psi_{\theta}= \delta \cos {\theta} \, \Psi_{\theta} , \quad  H_0 \widetilde{\Psi}_{\theta}= \delta \cos {\theta} \, \Psi_{\theta},
\end{align}
where $\theta$ controls the magnitude of  the eigenvalues while $\mu$ reflects the translational invariance of $H_0$. The states (\ref{eigenh}) have a definite parity with respect to $PT$ as long as both $\theta$ and $\mu$ are  simultaneously real or imaginary. In the latter case, they represent scattering states (linear combination of plane waves) with the eigenvalues $\lambda=\delta \cosh|\theta|$. The  scattering states corresponding to $-\lambda$ can be obtained by multiplication of the former ones by $\sigma_2$ as there holds $\{H_0,\sigma_2\}=0$.

Let us fix the following eigenstate and the associated Jordan state of $H_0$, 
\begin{align}\label{fpset1}
\Psi=\alpha\Psi_{\theta}+i\beta\widetilde{\Psi}_{\theta}, \quad \chi^{(1)}=\frac{\partial\Psi}{\partial \theta}+\nu\Psi_{\theta}+i\eta \widetilde{\Psi}_{\theta}.
\end{align}
The coefficients $\alpha$, $\beta$, $\nu$ and $\eta$ have to be fixed such that definite $PT$-parity of the states is guaranteed. For real $\theta$ and $\mu$, all $\alpha$, $\beta$, $\nu$ and $\eta$ have to be real as well. When $\theta$ and $\mu$ are purely imaginary, we have to take $\alpha$ and $\nu$ real  and $\beta$ and $\eta$ purely imaginary.

\begin{figure}[h!]
\centering
\includegraphics[scale=1]{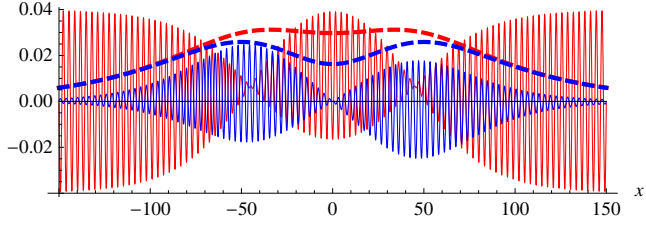}
\caption{(color online) Real (solid red) and imaginary (solid blue) part of the fluctuations $\delta n$ of the refractive index $n(x)$, $\delta n=n_0-n(x)$ (notice that real part is even while imaginary part is odd with respect to $PT$.). Absolute values of the upper (dashed red) and lower (dashed blue) components of $\widetilde{\Psi}^{(1)}$. We fixed $\rho(x)=\sigma(x)$, $\theta=\pi/2$, $\delta=0.02$, $\mu=0.5$, $\eta=1$, $\alpha=1$, $\beta=0$, $\nu=0$, $n_0=1$.\label{zeromodesgap}}
\label{polyhedra}
\end{figure}

Substituting (\ref{eigenh}) and (\ref{fpset1}) into either (\ref{V_1}) or (\ref{genrqn}) with $n=1$, the components of the potential term of the new Hamiltonian $H_1$ can be written as
\begin{align}\label{qr1}
q_1=\delta\left(-1+4\sin\theta\frac{(\alpha\cosh(x\delta\sin\theta+i\mu)+i\beta\sinh(x\delta\sin\theta+i\mu))^2}{D}\right),\\
r_1=\delta\left(-1-4\sin\theta\frac{(\alpha\cosh(x\delta\sin\theta+i\theta+i\mu)+i\beta\sinh(x\delta\sin\theta+i\theta+i\mu))^2}{D}\right),
\end{align}
where
\begin{align}
D=2\alpha\beta\cosh (2x\delta\sin\theta+i\theta+2i\mu)-(\alpha^2-\beta^2)i\sinh(2x\sin\theta+i\theta+2i\mu)\\+(\alpha^2+\beta^2+2\alpha\eta-2\beta\nu)\sin\theta-ix(\alpha^2+\beta^2)\delta\sin 2\theta.
\end{align}
The Hamiltonian $H_1$ possesses a \textit{bounded} zero mode provided that $\theta=\pi/2+k\pi$, where  $k$ is an integer. For other values of $\theta$, the zero modes are exponentially expanding. Fixing $\theta=\pi/2$, the zero modes of $H_1$ {\bf are $\Psi^{(1)}=L\chi^{(2)}_{\pi/2}$} and $\widetilde{\Psi}^{(1)}=L\widetilde{\Psi}_{\pi/2}$. The state $\Psi^{(1)}$ expands exponentially for $|x|\rightarrow\infty$ whereas $\widetilde{\Psi}^{(1)}$ is exponentially decaying. Its explicit form is
\begin{equation}\label{BS}
 \tilde{\Psi}^{(1)}=\left(\begin{array}{c}
\frac{2\alpha\delta(\alpha\cosh(x\delta+i\mu)+i\beta\sinh(x\delta+i\mu))}{\alpha^2+\beta^2+2\alpha\eta-2\beta\nu+(\alpha^2-\beta^2)\cosh(2x\delta+2i\mu)+2i\alpha\beta\sinh(2x\delta+2i\mu)}\\
\frac{2\alpha\delta(\beta\cosh(x\delta+i\mu)-i\alpha\sinh(x\delta+i\mu))}{\alpha^2+\beta^2+2\alpha\eta-2\beta\nu+(\alpha^2-\beta^2)\cosh(2x\delta+2i\mu)+2i\alpha\beta\sinh(2x\delta+2i\mu)}
\end{array}\right)\, .
\end{equation}
Substituting (\ref{qr1}) into (\ref{hde}) and utilizing the definition of $n(x)$ below (\ref{DE}), we can find the explicit form of the refractive index which is presented in the Fig.  (\ref{zeromodesgap}) together with (\ref{BS}).

\subsection{Power law decay of the electric field}
\begin{figure}[h!]
\centering
\includegraphics[scale=0.8]{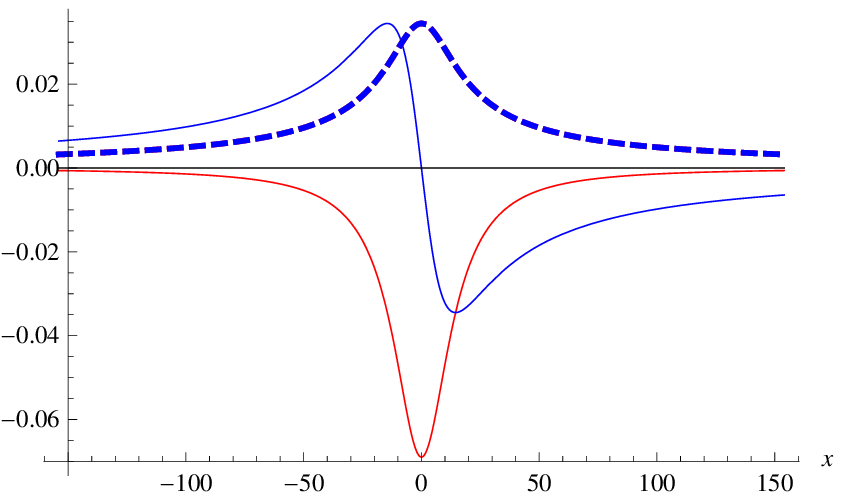} \includegraphics[scale=0.8]{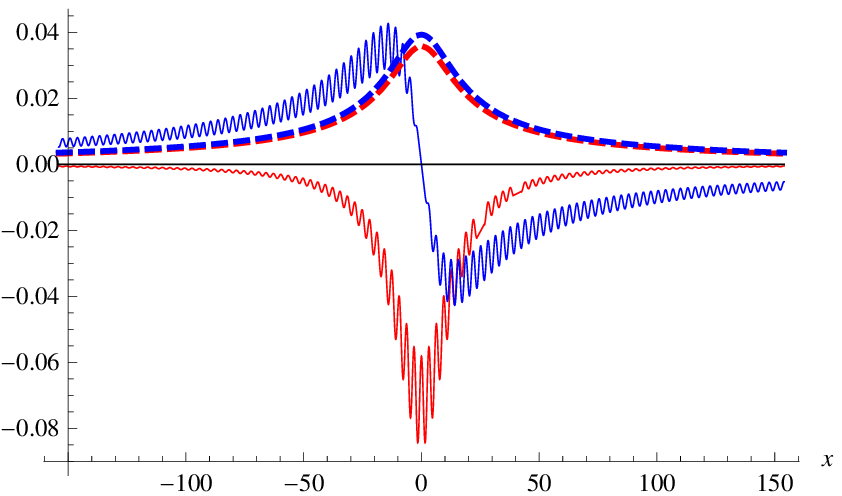} 
\caption{(color online) Real (solid red) and imaginary (solid blue) part of the fluctuations $\delta n$ of the refractive index $n(x)$, $\delta n=n_0-n(x)$. Absolute values of the upper (dashed red) and lower (dashed blue) components of $\widetilde{\Psi}^{(1)}$. The two components of $\widetilde{\Psi}^{(1)}$ coincide for $\alpha=-\beta$. We fixed $\alpha=-1$, $\beta=1$, $\nu=18$, $\eta=11$, $n_0=1$ in the left figure while we changed $\beta=1.1$ in the right figure.\label{delta0}}
\label{polyhedra}
\end{figure}

When $H_0$ lacks the mass term, $\delta=0$, the parametrization (\ref{eigenh}) is not suitable. Instead, we fix the eigenfunctions in the following manner
\begin{equation}
\Psi_\lambda=(1,0)^t e^{i \lambda x},\quad \widetilde{\Psi}_\lambda=(0,1)^t e^{-i \lambda x}
\end{equation}
\begin{equation}
H_0 \Psi_{\lambda}= \lambda \, \Psi_{\lambda} , \quad  H_0 \widetilde{\Psi}_{\lambda}= \lambda \, \widetilde{\Psi}_{\lambda}.
\end{equation}
We fix the kernel of the intertwining operator $L_1$ as
\begin{equation}
 \Psi=\alpha\Psi_0+\beta\widetilde{\Psi}_0,\quad \chi^{(1)}=\alpha \chi_0^{(1)}+\beta\widetilde{\chi}_0^{(1)}+\nu\Psi_0+\eta\widetilde{\Psi}_0,
\end{equation}
i.e. $L_1\Psi=L_1\chi^{(1)}=0$. $\Psi_0$ and $\widetilde{\Psi}_0$ are zero modes of $H_0$. The associated Jordan states are 
$\chi_0^{(1)}=(ix,0)^t e^{i \lambda x}$ and $\widetilde{\chi}_0^{(1)}=(0,-ix)^t e^{-i \lambda x}$ and satisfy $H_0\chi_0^{(1)}=\Psi_0,$ $H_0\widetilde{\chi}_0^{(1)}=\widetilde{\Psi}_0.$ Here, the states have definite parity provided that the coefficients $\alpha$, $\beta$, $\nu$ and $\eta$ are all either purely real or purely imaginary.
Then potential term of the Hamiltonian $H_1$ reads
\begin{equation}
 q=\frac{2\alpha^2 e^{2i \lambda x}}{\alpha \eta-\beta \nu-2ix\alpha\beta},\quad r=-\frac{2\beta^2 e^{-2i \lambda x}}{\alpha \eta-\beta \nu-2ix\alpha\beta}
\end{equation}
From the two zero modes of $H_1$ only one is bounded and reads
\begin{equation}
 \Psi_1=\frac{i\beta}{\alpha \eta-\beta \nu-2ix\alpha\beta}\left(\begin{array}{c}\alpha\\-\beta\end{array}\right).
\end{equation}
Let us notice that upper and lower components of $\Psi_1$ coincide for $\beta=-\alpha$. The plot of the modulus of the components of $\Psi_1$ together with the corresponding fluctuations of the refractive index are in Fig. \ref{delta0}.

\section{Discussion and Outlook}
In the current article, we focused on the analysis of the optical systems described in (\ref{DE}) with $PT$-symmetric potential that simulates a balanced gain and loss of signal in the optical setting. In order to construct exactly solvable models with these properties, we presented  the extension of the confluent Crum-Darboux transformation  mechanism for the one-dimensional Dirac equation.  The construction of the intertwining operator between Dirac Hamitonians as well as chains of Crum-Darboux transformations were discussed in detail. We presented the explicit form of the higher-order intertwining operator and the associated new Hamiltonian, see (\ref{intertn}) and (\ref{htilde}), which can be used not only for confluent Crum-Darboux transformation but also for usual and combined ones. The machinery was illustrated in the preceding section, where the transformation was utilized for construction of exactly solvable Dirac Hamiltonians that corresponded to mild, complex fluctuations of the refractive index. We presented regimes where intensity of electric field is localized and decays either exponentially or as $\sim 1/x$. 

It is worth noticing that in \cite{AxelCDirac}, the confluent Crum-Darboux transformation coincided with  intertwining operator between two Schr\"odinger-type operators, obtained as the squares of the first order Dirac Hamiltonians. Hence, the construction presented there was limited to Dirac Hamiltonians with pseudo-scalar potentials. In our case, the intertwining operator is constructed such that it intertwines directly the Dirac Hamiltonians and the potential term of the Hamiltonian is not restricted. 
In \cite{longhiPRL}, a optical realization of $PT$-symmetric relativistic systems was discussed such that a more general ansatz for the electric field in Bragg grating was used; the functions $u$ and $v$ were allowed to depend on time. Then (\ref{DE}) acquires the form of time-dependent Dirac equation, see the footnote below (\ref{DE}). The confluent Darboux transformation was utilized in \cite{CJPCSUSY} for the analysis of the optical systems with invisible defects. It might be interesting to analyze similar systems described by Hamiltonians of the Dirac type with the use of our current results.
 
The framework presented here is based on the time-independent, one-dimensional Dirac Hamiltonian. The scheme presented here could be extended to the analysis of the time-dependent systems that are useful in the context of solutions of non-linear integrable equations, e.g. equations of the integrable Ablowitz-Kaup-Newell-Segur hierarchy \cite{akns}. In this context, confluent Crum-Darboux transformations of Schr\"odinger Hamiltonian were utilized to find $PT$-symmetric multi-soliton solutions with non-trivial behavior \cite{Correa:2016zrb}.

The current results can be also employed for the analysis of the Schr\"odinger operators with matrix potentials. In \cite{TranspDirac}, Crum-Darboux transformation for Dirac Hamiltonian was utilized for construction of Schr\"odinger operators with transparent matrix potential. It is worth noticing that these operators were analyzed recently by in different manner in \cite{Sokolov1}, \cite{Sokolov2}. However, this analysis would go beyond the scope of the present article.   
\vspace{2mm}

{\bf Acknowledgments} 
FC is 
supported by the Alexander von Humboldt Foundation. 
VJ was supported by the GA\v CR Grant No.15-07674Y. 
FC thanks Nuclear Physics Institute of Czech Republic, 
where a part of this work was done, for hospitality.

\underline{}

\end{document}